\documentclass{article}%
\usepackage{amssymb}
\usepackage{makeidx}
\usepackage{amsfonts}
\usepackage{amsmath}
\usepackage{color}
\usepackage{harvard}
\usepackage{framed}
\usepackage{graphicx}%
\providecommand{\U}[1]{\protect\rule{.1in}{.1in}}

\let\cite=\citeasnoun
\begin{document}

\begin{center}%
\begin{tabular}
[c]{c}%
{\Large Fully Flexible Views: Theory and Practice}\footnote{This article
appears as Meucci A., 2008, \textit{Fully Flexible Views: Theory and
Practice}, Risk, \textbf{21} (10) 97-102}\\
\\
{\large Attilio Meucci}\footnote{The author is grateful to Paul Glasserman,
Sridhar Gollamudi, Ninghui Liu and an anonymous referee for their helpful
feedback; and to aorda.com for providing Portfolio Safeguard to benchmark some
numerical computations}\\
attilio\_meucci@symmys.com\\
\\
\\
this version:\ December 13 2010\\
latest version available at \texttt{http://ssrn.com/abstract=1213325}%
\end{tabular}

\end{center}

\bigskip

\begin{center}
\textbf{Abstract}
\end{center}

\noindent We propose a unified methodology to input non-linear views from any
number of users in fully general non-normal markets, and perform, among
others, stress-testing, scenario analysis, and ranking allocation. We walk the
reader through the theory and we detail an extremely efficient algorithm to
easily implement this methodology under fully general assumptions. As it turns
out, no repricing is ever necessary, hence the methodology can be readily
applied to books with complex derivatives. We also present an analytical
solution, useful for benchmarking, which per se generalizes notable previous
results. Code illustrating this methodology in practice is available
at\newline\texttt{http://www.mathworks.com/matlabcentral/fileexchange/21307}.

\bigskip

\bigskip

\bigskip\bigskip\newpage

\section{Introduction}

Scenario analysis allows the practitioner to explore the implications on a
given portfolio of a set of subjective views on possible market realizations,
see e.g. \cite{MinaXiao01}. The pathbreaking approach pioneered by
\cite{BlackLitt90} (BL\ in the sequel) generalizes scenario analysis, by
adding uncertainty on the views and on the reference risk model. Further
generalizations have been proposed in recent years. \cite{Qian01} provide a
framework to stress-test volatilities and correlations in addition to
expectations. \cite{Pezier07} processes partial views on expectations and
covariances based on least discrimination. \cite{Meucci08f} extends the above
models to act on risk factors instead of returns, and thus covers highly
non-linear derivative markets and views on external factors that influence the
p\&l only statistically.

In the above techniques, the reference distribution of the risk factors is
normal. The COP in \cite{Meucci06b} explores non-normal markets, but
correlation stress-testing and non-linear views are not allowed. Furthermore,
the COP\ relies on ad-hoc manipulations.

Here we present the entropy pooling approach (EP in the sequel) which fully
generalizes the above and related techniques. The inputs are an arbitrary
market model, which we call "prior", and fully general views or stress-tests
on that market. The output is a distribution, which we call "posterior", that
incorporates all the inputs and can be used for risk management and portfolio optimization.

To obtain the posterior, we interpret the views as statements that distort the
prior distribution, in such a way that the least possible amount of spurious
structure is imposed. The natural index for the structure of a distribution is
its entropy. Therefore we define the posterior distribution as the one that
minimizes the entropy relative to the prior. Then by opinion pooling we assign
different confidence levels to different views and users.

Among others, the EP handles non-normal markets; views on non-linear
combinations of risk factors that impact the p\&l directly or only
statistically through correlations; views on expectations, but also medians,
to handle fat tails; views on volatilities, correlations, tail behaviors,
etc.; lax views, such as ranking, on all of the above, thereby generalizing
\cite{AlmgrenChriss06}; inputs from multiple users and multiple confidence
levels for different views.

Furthermore, in its most general implementation the reference model is
represented by Monte Carlo simulations, and the posterior which incorporates
all the inputs is represented by the \textit{same} simulations with new
probabilities.\ Hence the most complex securities can be handled without
costly repricing.

In Section \ref{SecTHaoifgasodg} we introduce the EP theoretical framework. In
Section \ref{SecImapea copy(1)} we present an analytical formula, which
generalizes the previous results and provides a benchmark for the numerical
implementation. In Section \ref{SecImapea} we discuss the numerical routine to
implement the EP in full generality. In Section \ref{Exzagwga} we illustrate a
case study: option trading in a non-normal environment with non-linear and
ranking views on realized volatility, implied volatility and external macro
factors. In Section \ref{SeccdCndogd} we conclude, comparing the EP to other
related techniques. Fully documented code for this and other case studies,
such as portfolios from ranking, can be downloaded at MATLAB Central File Exchange.

\section{The entropy pooling approach\label{SecTHaoifgasodg}}

We consider a book driven by an $N$-dimensional vector of risk factors
$\mathbf{X}$. In other words, denoting by $t$ the current time, by
$\mathcal{I}_{t}$ the information currently available, and by $\tau$ the time
to the investment horizon, there exists a deterministic function $P$ that maps
the realizations of $\mathbf{X}$ and the information $\mathcal{I}_{t}$ into
the price $P_{t+\tau}$ of each security in the book at the horizon:%
\begin{equation}
P_{t+\tau}\equiv P\left(  \mathbf{X},\mathcal{I}_{t}\right)
\text{.\label{PRoixag}}%
\end{equation}
This framework is completely general. For instance, in a book of options
$\mathbf{X}$ can represent the changes in all the underlyings and implied
volatilities: in this case $\left(  \ref{PRoixag}\right)  $ is approximated by
a second-order Taylor expansion whose coefficients are the "deltas", "vegas",
"gammas", "vannas", "volgas", etc. Also, $\mathbf{X}$ can represent a set of
risk factors behind a computationally expensive full Monte-Carlo pricing
function, such as interest rate values at different monitoring times for
mortgage derivatives. Furthermore, $\mathbf{X}$ can be augmented with a set of
external risk factors that do not feed directly the pricing function $\left(
\ref{PRoixag}\right)  $, but that still influence the p\&l statistically
through correlation. We explore a detailed example in these directions in
Section \ref{Exzagwga}. In any case, we emphasize that $\mathbf{X}$ can be,
but by no means is restricted to, returns on a set of securities.

\textbf{The reference model}\newline We assume the existence of a risk model,
i.e. a model for the joint distribution of the risk factors, as represented by
its probability density function (pdf)%
\begin{equation}
\mathbf{X}\sim f_{\mathbf{X}}\text{.} \label{MAara}%
\end{equation}
In BL, this is the "prior" factor distribution. More in general, this is a
model that risk managers use to perform risk analyses, such as the computation
of the volatility, tracking error, VaR, expected shortfall of a portfolio,
along with the contributions to such measures from the different sources of
risk. Portfolio managers and traders on the other hand use this model to
optimize their positions. They specify a subjective index of satisfaction
$\mathcal{S}$, such as the mean-(C)VaR trade-off, or the certainty equivalent
stemming from a utility function, or a spectral measure, etc., see examples in
\cite{Meucci05}. Satisfaction depends both on the market distribution
$f_{\mathbf{X}}$ through the prices $\left(  \ref{PRoixag}\right)  $ and on
the positions in the book, represented by a vector $\mathbf{w}$. Then the
optimal book $\mathbf{w}^{\ast}$ is defined as%
\begin{equation}
\mathbf{w}^{\ast}\equiv\operatorname*{argmax}_{\mathbf{w}\in\mathcal{C}%
}\left\{  \mathcal{S}\left(  \mathbf{w};f_{\mathbf{X}}\right)  \right\}
\text{,\label{asgdsedhid}}%
\end{equation}
where $\mathcal{C}$ is a given set of investment constraints. The reference
model $\left(  \ref{MAara}\right)  $ can be estimated from historical
analysis, or calibrated to current market observables, see \cite{Meucci08f}.

\textbf{The views}\newline In the most general case, the user expresses views
on generic functions of the market $g_{1}\left(  \mathbf{X}\right)
,\ldots,g_{K}\left(  \mathbf{X}\right)  $. These functions constitute a
$K$-dimensional random variable whose joint distribution is implied by the
reference model $\left(  \ref{MAara}\right)  $:%
\begin{equation}
\mathbf{V}\equiv\mathbf{g}\left(  \mathbf{X}\right)  \sim f_{\mathbf{V}%
}\text{.} \label{MAfa}%
\end{equation}
We emphasize that, unlike in BL, in EP we do not assume that the functions
$g_{k}$ be linear. Notice that, as a special case, one can express views also
on the securities values $\left(  \ref{PRoixag}\right)  $.

The views, or the stress-tests, are statements on the variables $\left(
\ref{MAfa}\right)  $ which can clash with the reference model. In a stochastic
environment, this means statements on their distribution. Therefore, the most
detailed possible view specification is a complete, subjective joint
distribution for those variables:%
\begin{equation}
\mathbf{V}\sim\widetilde{f}_{\mathbf{V}}\neq f_{\mathbf{V}}%
\text{.\label{MAfa copy(1)}}%
\end{equation}
However, views in general are statements on only select features of the
distribution of $\mathbf{V}$.

\begin{itemize}
\item The classical views a-la BL are statements on $\widetilde{\mathbb{E}%
}\left\{  V_{k}\right\}  $, the expectations of each of the $V_{k}$'s
according to the new distribution $\widetilde{f}_{\mathbf{V}}$. Since for
distributions such as stable distributions the expectation is not defined, in
EP\ we consider views on a more general location measure $\widetilde
{m}\left\{  V_{k}\right\}  $, which can be the expectation or the median. The
views are then set as%
\begin{equation}
\widetilde{m}\left\{  V_{k}\right\}  \gtreqqless m_{k},\quad k=1,\ldots
,K\text{,} \label{BUllwivas}%
\end{equation}
The values $m_{k}$ can be determined exogenously. If the user has only
qualitative views, it is convenient to set as in \cite{Meucci08a}%
\begin{equation}
m_{k}\equiv m\left\{  V_{k}\right\}  +\varkappa\sigma\left\{  V_{k}\right\}
\text{.\label{PSfadgdg}}%
\end{equation}
In this expression $\sigma$ is a measure of volatility in the reference model,
such as the standard deviation or, in fat-tailed markets with infinite
variance, the interquartile range; and $\varkappa$ is an ad-hoc multiplier,
such as $-2$, $-1$, $1$, and $2$ for "very bearish", "bearish", "bullish" and
"very bullish" respectively.

\item The generalized BL views $\left(  \ref{BUllwivas}\right)  $ are not
necessarily expressed as equality constraint:\ EP can process views expressed
as inequalities. In particular, EP can process ordering information, frequent
in stock and bond management:%
\begin{equation}
\widetilde{m}\left\{  V_{1}\right\}  \geq\widetilde{m}\left\{  V_{2}\right\}
\geq\cdots\geq\widetilde{m}\left\{  V_{K}\right\}  \text{.}
\label{SyoafkaRand}%
\end{equation}

\item Views can be expressed on the volatilities. A convenient formulation
reads:%
\begin{equation}
\widetilde{\sigma}\left\{  V_{k}\right\}  \gtreqqless\varkappa\sigma\left\{
V_{k}\right\}  ,\quad k=1,\ldots,K\text{.} \label{tasbVOslafs}%
\end{equation}

\item Correlation stress-tests are also views. Convenient specifications for
the correlation matrix $\widetilde{\mathbb{C}}\left\{  \mathbf{V}\right\}  $
are the homogeneous shrinkage%
\begin{equation}
\widetilde{\mathbb{C}}\left\{  \mathbf{V}\right\}  \equiv\rho_{1}%
\mathbf{I}+\rho_{2}\mathbb{C}\left\{  \mathbf{V}\right\}  +\rho_{3}%
\mathbf{11}^{\prime}\text{,\label{COtdefs}}%
\end{equation}
where $0\leq\rho_{1},\rho_{2},\rho_{3}<1$, $\rho_{1}+\rho_{2}+\rho_{3}\equiv
1$, $\mathbf{I}$ is the identity matrix and $\mathbf{1}$ is a vector of ones.
For different structures see e.g. \cite{BrigoMercurio01}.

\item The user can input views on the lower (upper) tail behavior, as
represented e.g. by $\widetilde{Q}_{V}\left(  u\right)  $, the quantile of
$V_{k}$ according to the new distribution $\widetilde{f}_{\mathbf{V}}$, where
the tail level $u$ is close to zero (one). A convenient specification is%
\begin{equation}
\widetilde{Q}_{V}\left(  u\right)  \gtreqqless Q_{V}\left(  u\right)
\text{,\label{GTIsalzidgd copy(1)}}%
\end{equation}
where $Q_{V}$ is the reference quantile induced by $f_{\mathbf{V}}$, or
alternatively benchmark quantiles such as the normal or the Student $t$.

\item Lower (upper) tail codependence, as represented by $\widetilde
{C}_{\mathbf{V}}\left(  \mathbf{u}\right)  $, the cdf of the copula of
$\mathbf{V}$ at joint threshold levels $\mathbf{u}$ close to zero (one). A
convenient specification reads%
\begin{equation}
\widetilde{C}_{\mathbf{V}}\left(  \mathbf{u}\right)  \gtreqqless\varkappa
C_{\mathbf{V}}\left(  \mathbf{u}\right)  \text{,\label{GTIsalzidgd}}%
\end{equation}
where $C_{\mathbf{V}}$ is the reference copula cdf induced by $f_{\mathbf{V}}
$, or alternatively benchmark copula cdf's such as normal or Student $t$.
\end{itemize}

The above is a very partial list of all the possible features on which the
user can wish to express views, and which can be handled by the EP.

\textbf{The posterior}\newline The posterior distribution should satisfy the
views without adding additional structure and should be as close as possible
to the reference model $\left(  \ref{MAara}\right)  $.

The relative entropy between a generic distribution $\widetilde{f}%
_{\mathbf{X}}$ and a reference distribution $f_{\mathbf{X}}$
\begin{equation}
\mathcal{E}\left(  \widetilde{f}_{\mathbf{X}},f_{\mathbf{X}}\right)
\equiv\int\widetilde{f}_{\mathbf{X}}\left(  \mathbf{x}\right)  \left[
\ln\widetilde{f}_{\mathbf{X}}\left(  \mathbf{x}\right)  -\ln f_{\mathbf{X}%
}\left(  \mathbf{x}\right)  \right]  d\mathbf{x}\text{.\label{SVIhjasJKL}}%
\end{equation}
is a natural measure of the amount of structure in $\widetilde{f}_{\mathbf{X}%
}$; furthermore, it also measures how distorted $\widetilde{f}_{\mathbf{X}}$
is with respect to $f_{\mathbf{X}}$. Indeed, if the two distributions
coincide, relative entropy is zero; by imposing constraints on $\widetilde
{f}_{\mathbf{X}}$ this distribution departs from $f_{\mathbf{X}}$ and relative
entropy increases.

Therefore, we define the posterior market distribution as%
\begin{equation}
\widetilde{f}_{\mathbf{X}}\equiv\operatorname*{argmin}_{f\in\mathbb{V}%
}\left\{  \mathcal{E}\left(  f,f_{\mathbf{X}}\right)  \right\}
\text{,\label{FLSKnadd copy(1)}}%
\end{equation}
where $f\in\mathbb{V}$ stands for all the distributions consistent with the
views statements such as $\left(  \ref{BUllwivas}\right)  $-$\left(
\ref{GTIsalzidgd}\right)  $.

Entropy minimization is widely applied in physics and statistics, see
\cite{CoverThomas06}. For applications to finance, see e.g.
\cite{Avellaneda99}, \cite{DAmFusTa03}, \cite{ContTankov07} and
\cite{Pezier07}. In our context, entropy minimization is even more natural, as
it generalizes Bayesian updating, see \cite{CatichaGiffin06}.

\textbf{The confidence}\newline One last step is required:\ the posterior
$\widetilde{f}_{\mathbf{X}}$ follows by assuming that the practitioner has
full confidence in his statements. If the confidence is less than full, the
posterior distribution of the factors must shrink towards the reference factor
distribution. This is easily achieved as in \cite{Meucci06b} by
opinion-pooling the reference model and the full-confidence posterior:%
\begin{equation}
\widetilde{f}_{\mathbf{X}}^{c}\equiv\left(  1-c\right)  f_{\mathbf{X}%
}+c\widetilde{f}_{\mathbf{X}}\text{.\label{Sateg}}%
\end{equation}
The pooling parameter $c\in\left[  0,1\right]  $ represents the confidence
level in the views: in the extreme case when the confidence is total, the
full-confidence posterior is recovered; on the other hand, in the absence of
confidence, the reference risk model is recovered.

Opinion pooling becomes very useful in a multi-manager context. Indeed,
consider $S$ users that input their separate views on (possibly, but not
necessarily) different functions of the market. As in $\left(
\ref{FLSKnadd copy(1)}\right)  $, we obtain $S$ full-confidence posterior
distributions $\widetilde{f}_{\mathbf{X}}^{\left(  s\right)  }$,
$s=1,\ldots,S$. Then the posterior distribution results naturally as the
confidence-weighted average of the individual full-confidence posteriors:%
\begin{equation}
\widetilde{f}_{\mathbf{X}}^{\mathbf{c}}\equiv\sum_{s=1}^{S}c_{s}\widetilde
{f}_{\mathbf{X}}^{\left(  s\right)  }\text{.\label{Mulaigte}}%
\end{equation}
These confidence levels can be linked naturally to the track-record of the
respective manager, i.e. the $s$-th confidence $c_{s}$ can be set as an
increasing function of the number of past views, i.e. seniority, and of the
correlation of these views with the actual market realization, in the same
spirit as the "skill" measure in \cite{GrinoldKahn99}.

The definitions $\left(  \ref{Sateg}\right)  $-$\left(  \ref{Mulaigte}\right)
$ follow from a probabilistic interpretation of the confidence: one can easily
specify different confidence levels for the different views of the same user
and integrate these within a multi-user context. As it turns out, this amounts
to specifying a probability measure on the power set of the views: we discuss
these simple rules in detail in Appendix \ref{AppPRvab}.

We emphasize that, unlike in BL, in EP\ the confidence in the views $\left(
\ref{Sateg}\right)  $ and the views on volatility $\left(  \ref{tasbVOslafs}%
\right)  $ are modeled separately: indeed, being sure about future volatility
and being uncertain about future market realizations are two very different issues.

\textbf{Limit cases}\newline If the practitioner has no views, i.e.
$\mathbb{V}$\ is the empty set in $\left(  \ref{FLSKnadd copy(1)}\right)  $,
then the confidence-weighted posterior distribution equals the reference model
$f_{\mathbf{X}}$.

On the other extreme, if the views fully specify a joint distribution $\left(
\ref{MAfa copy(1)}\right)  $ the minimization $\left(  \ref{FLSKnadd copy(1)}%
\right)  $ is not necessary. Indeed, consistently with the principle of
minimum discrimination information, the full-confidence posterior follows from
its conditional-marginal decomposition:%
\begin{equation}
\widetilde{f}_{\mathbf{X}}\left(  \mathbf{x}\right)  \equiv\int f_{\mathbf{X}%
|\mathbf{v}}\left(  \mathbf{x}\right)  \widetilde{f}_{\mathbf{V}}\left(
\mathbf{v}\right)  d\mathbf{v}\text{.\label{FUlagsgas}}%
\end{equation}
In particular, this is the case in scenario analysis, where the user
associates full probability to one single scenario $\mathbf{g}\left(
\mathbf{X}\right)  \equiv\widetilde{\mathbf{v}}$: the views are represented
with a Dirac delta centered on the scenario $\widetilde{f}_{\mathbf{V}}\left(
\mathbf{v}\right)  \equiv\delta\left(  \mathbf{v}-\widetilde{\mathbf{v}%
}\right)  $, which, substituted in $\left(  \ref{FUlagsgas}\right)  $, yields
$\widetilde{f}_{\mathbf{X}}\equiv f_{\mathbf{X}|\widetilde{\mathbf{v}}}$. In
words, the full-confidence posterior distribution is simply the reference
distribution, conditioned on $\mathbf{g}\left(  \mathbf{X}\right)  $ assuming
the scenario values $\widetilde{\mathbf{v}}$. Therefore, EP includes
full-distribution specification and standard scenario analysis as special cases.

\section{An analytical formula\label{SecImapea copy(1)}}

Consider as in BL a normal reference model%
\begin{equation}
\mathbf{X}\sim\operatorname{N}\left(  \mathbf{\mu},\mathbf{\Sigma}\right)
\text{.\label{Signaosgs}}%
\end{equation}
Consider views on the expectations of arbitrary linear combinations
$\mathbf{QX}$ and on the covariances of arbitrary, potentially different,
linear combinations $\mathbf{GX}$%
\begin{equation}
\mathbb{V}:\left\{
\begin{tabular}
[c]{l}%
$\widetilde{\mathbb{E}}\left\{  \mathbf{QX}\right\}  \equiv\widetilde
{\mathbf{\mu}}_{\mathbf{Q}}$\\
$\widetilde{\mathbb{C}}ov\left\{  \mathbf{GX}\right\}  \equiv\widetilde
{\mathbf{\Sigma}}_{\mathbf{G}}$,
\end{tabular}
\right.  \label{Coaidsgna}%
\end{equation}
where $\mathbf{Q}$, $\mathbf{G}$, $\widetilde{\mathbf{\Sigma}}_{\mathbf{G}}$
and $\widetilde{\mathbf{\mu}}_{\mathbf{Q}}$ are conformable matrices/vector.

As we show in Appendix \ref{Appfasfppdfd}, the full-confidence posterior
distribution $\left(  \ref{FLSKnadd copy(1)}\right)  $ is normal:%
\begin{equation}
\mathbf{X}\sim\operatorname{N}\left(  \widetilde{\mathbf{\mu}},\widetilde
{\mathbf{\Sigma}}\right)  \text{,\label{POsterw}}%
\end{equation}
where%
\begin{align}
\widetilde{\mathbf{\mu}}  &  \equiv\mathbf{\mu}+\mathbf{\Sigma Q}^{\prime
}\left(  \mathbf{Q\Sigma Q}^{\prime}\right)  ^{-1}\left(  \widetilde
{\mathbf{\mu}}_{\mathbf{Q}}-\mathbf{Q\mu}\right)  \text{,\label{MMuasfas}}\\
\widetilde{\mathbf{\Sigma}}  &  \equiv\mathbf{\Sigma}+\mathbf{\Sigma
G}^{\prime}\left(  \left(  \mathbf{G\Sigma G}^{\prime}\right)  ^{-1}%
\widetilde{\mathbf{\Sigma}}\mathbf{_{\mathbf{G}}}\left(  \mathbf{G\Sigma
G}^{\prime}\right)  ^{-1}-\left(  \mathbf{G\Sigma G}^{\prime}\right)
^{-1}\right)  \mathbf{G\Sigma}\text{.} \label{BBuas}%
\end{align}
Then the confidence-weighted posterior distribution $\left(  \ref{Sateg}%
\right)  $ is a normal mixture:%
\begin{equation}%
\begin{tabular}
[c]{llll}
&  & $\operatorname{N}\left(  \mathbf{\mu},\mathbf{\Sigma}\right)  $ &
(probability: $1-c$)\\
& $\nearrow$ &  & \\
$\mathbf{X}\sim$ &  &  & \\
& $\searrow$ &  & \\
&  & $\operatorname{N}\left(  \widetilde{\mathbf{\mu}},\widetilde
{\mathbf{\Sigma}}\right)  $ & (probability: $c$)
\end{tabular}
\label{boiajasgfsd}%
\end{equation}
This distribution is suitable for instance to stress-test market crashes,
where\ high volatilities, high correlations and low expectations in
$\widetilde{\mathbf{\mu}},\widetilde{\mathbf{\Sigma}}$ are expected to occur
with probability $c\ll1$.

Formula $\left(  \ref{boiajasgfsd}\right)  $ generalizes results in
\cite{Pezier07}. Also, the special case of full-confidence $c\equiv1$ on only
one set of linear combinations $\mathbf{Q}\equiv\mathbf{G}$ yields the result
in \cite{Qian01}: this is not surprising, as the authors' approach is
equivalent to the decomposition $\left(  \ref{FUlagsgas}\right)  $. Finally,
the further specialization to null dispersion in the views $\widetilde
{\mathbf{\Sigma}}\mathbf{_{\mathbf{G}}}\rightarrow\mathbf{0}$, yields scenario
analysis as in \cite{Meucci05}, which in turn generalizes the standard
regression-based approach that appears e.g. in \cite{MinaXiao01}.

\section{Numerical implementation\label{SecImapea}}

Except for the special case in Section \ref{SecImapea copy(1)}, the EP cannot
be implemented analytically. However, the numerical implementation of the EP
in full generality is extremely simple and computationally efficient.

First, we represent the reference distribution $\left(  \ref{MAara}\right)  $
of the market $\mathbf{X}$ in terms of a $J\times N$ panel $\mathcal{X}$ of
simulations: the generic $j$-th row of $\mathcal{X}$ represents one in a very
large number of joint scenarios for the $N$ variables $\mathbf{X}$, whereas
the generic $n$-th column of $\mathcal{X}$ represents the marginal
distribution of the $n$-th factor $X_{n}$. With the scenarios we associate the
$J\times1$ vector of the respective probabilities $\mathbf{p}$, whose each
entry typically, but not necessarily, equals $1/J$, see \cite{GlassermanYu05}
for a variety of methods to determine $\mathbf{p}$.

We assume that each of the joint scenarios in $\mathcal{X}$ has been mapped
into the respective joint price scenarios for the $I$ securities in the market
considered by the user, by means of the potentially costly function $\left(
\ref{PRoixag}\right)  $, thereby generating a $J\times I$ panel of prices
$\mathcal{P}$. The panel of the security prices $\mathcal{P}$, along with the
respective probabilities $\mathbf{p}$, is then analyzed for risk management
purposes, or it is fed into an optimization algorithm to perform the asset
allocation step $\left(  \ref{asgdsedhid}\right)  $.

The user expresses views on generic non-linear functions of the market
$\left(  \ref{MAfa}\right)  $. Their distribution as implied by the reference
model is readily represented by the $J\times K$ panel $\mathcal{V}$ defined
entry-wise as follows:%
\begin{equation}
\mathcal{V}_{j,k}\equiv g_{k}\left(  \mathcal{X}_{j,1},\ldots,\mathcal{X}%
_{j,N}\right)  \text{,\label{VScene}}%
\end{equation}

To represent the posterior distribution of the market that includes the views,
instead of generating new simulations, we use the \textit{same} scenarios with
different probabilities $\widetilde{\mathbf{p}}$. Then, as we show in Appendix
\ref{APpdgiasdpg}, general views such as $\left(  \ref{BUllwivas}\right)
$-$\left(  \ref{GTIsalzidgd}\right)  $ can be written as a set of linear
constraints on the new, yet to be determined, probabilities%
\begin{equation}
\underline{\mathbf{a}}\leq\mathbf{A}\widetilde{\mathbf{p}}\leq\overline
{\mathbf{a}}\text{,} \label{Cliagad}%
\end{equation}
where $\mathbf{A}$, $\underline{\mathbf{a}}$ and $\overline{\mathbf{a}}$ are
simple expressions of the panel $\left(  \ref{VScene}\right)  $. For instance,
for standard views on expectations $\mathbf{A}\equiv\mathcal{V}^{\prime}$ and
$\underline{\mathbf{a}}\equiv\overline{\mathbf{a}}$ quantify the views.%

\begin{figure}[h]%
\centering
\includegraphics[
height=201.5pt,
width=302.25pt
]%
{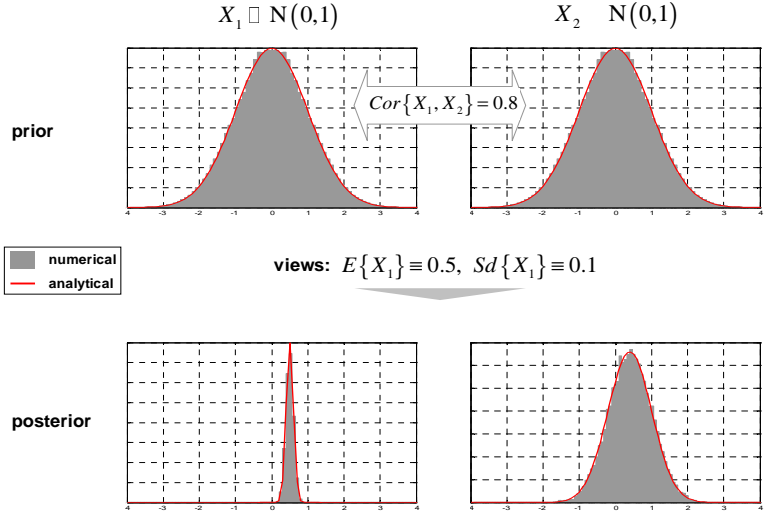}%
\caption{Entropy pooling:\ numerical approach matches analytical solution}%
\label{FigPOsaSFaczc}%
\end{figure}
Furthermore, the relative entropy $\left(  \ref{SVIhjasJKL}\right)  $ becomes
its discrete counterpart%
\begin{equation}
\mathcal{E}\left(  \widetilde{\mathbf{p}},\mathbf{p}\right)  \equiv\sum
_{j=1}^{J}\widetilde{p}_{j}\left[  \ln\left(  \widetilde{p}_{j}\right)
-\ln\left(  p_{j}\right)  \right]  \text{.} \label{Egnadogids}%
\end{equation}
Therefore, the full-confidence posterior distribution $\left(
\ref{FLSKnadd copy(1)}\right)  $ is defined as%
\begin{equation}
\widetilde{\mathbf{p}}\equiv\operatorname*{argmin}_{\underline{\mathbf{a}}%
\leq\mathbf{Af}\leq\overline{\mathbf{a}}}\left\{  \mathcal{E}\left(
\mathbf{f},\mathbf{p}\right)  \right\}  \text{.\label{acsggad}}%
\end{equation}
This optimization can be solved very efficiently: as we show in Appendix
\ref{AppSovEBter}, the dual formulation is a simple linearly constrained
convex program in a number of variables equal to the number of views, not the
number of Monte Carlo simulations, which can be kept large. Therefore we can
achieve an excellent accuracy even under extreme views, see Figure
\ref{FigPOsaSFaczc}.

Now it is immediate to compute the opinion-pooling, confidence-weighted
posterior $\left(  \ref{Sateg}\right)  $: this is represented by $\left(
\mathcal{X},\mathbf{p}_{c}\right)  $, the same simulations as for the
reference model, but with new probabilities%
\begin{equation}
\mathbf{p}_{c}\equiv\left(  1-c\right)  \mathbf{p}+c\widetilde{\mathbf{p}%
}\text{.\label{VOaggadf}}%
\end{equation}
A similar expression holds for the more general multi-user, multi-confidence
posterior discussed in Appendix \ref{AppPRvab}.

Since the posterior factor distribution is obtained by tweaking the relative
probabilities of the scenarios $\mathcal{X}$ without affecting the scenarios
themselves, the posterior distribution of the market prices is represented by
$\left(  \mathcal{P},\mathbf{p}_{c}\right)  $, the \textit{original} panel of
joint prices and the new probabilities. Hence\ no repricing is necessary to
process views and stress-tests.

\section{Case study: option trading\label{Exzagwga}}

As in \cite{Meucci08f}, we consider a trader of butterflies, defined as long
positions in one call and one put with the same strike, underlying, and time
to maturity. The price $P_{t+\tau}$ of the butterfly at the investment horizon
can be written in the format $\left(  \ref{PRoixag}\right)  $ as a
deterministic non-linear function of a set of risk factors and current
information. Indeed%
\begin{equation}
P_{t+\tau}=BS\left(  y_{t}e^{X_{y}},h\left(  y_{t}e^{X_{y}},\sigma
_{t}+X_{\sigma},K,T-\tau\right)  ;K,T-\tau,r\right)  \text{.} \label{PRasf}%
\end{equation}
In this expression $\tau$ is the investment horizon; $y_{t}$ is the current
value and $X_{y}\equiv\ln\left(  y_{t+\tau}/y_{t}\right)  $ is the log-change
of the underlying; $\sigma_{t}$ is the current value and $X_{\sigma}%
\equiv\sigma_{t+\tau}-\sigma_{t}$ is the change in ATM implied volatility;
$BS$ is the Black-Scholes formula%
\begin{equation}
BS\left(  y,\sigma;K,T,r\right)  \equiv y\left[  \Phi\left(  d_{1}\right)
-\Phi\left(  -d_{1}\right)  \right]  -Ke^{-rT}\left[  \Phi\left(
d_{2}\right)  -\Phi\left(  -d_{2}\right)  \right]  \text{,\label{PRicsnafg}}%
\end{equation}
where $\Phi$ is the standard normal cdf; $K$ is the strike; $T$ is the time to
expiry; $r$ is the risk-free rate; $d_{1}\equiv\left(  \ln\left(  y/K\right)
+\left(  r+\sigma^{2}/2\right)  T\right)  /\sigma\sqrt{T}$, $d_{2}\equiv
d_{1}-\sigma\sqrt{T}$; and $h$ is a skew/smile map%
\begin{equation}
h\left(  y,\sigma;K,T\right)  \equiv\sigma+\alpha\frac{\ln\left(  y/K\right)
}{\sqrt{T}}+\beta\left(  \frac{\ln\left(  y/K\right)  }{\sqrt{T}}\right)
^{2}\text{,}%
\end{equation}
for coefficients $\alpha$ and $\beta$ which depend on the underlying and are
fitted empirically, similarly to \cite{Malz97}. If the investment horizon
$\tau$ is short, a delta-gamma-vega approximation of $\left(  \ref{PRasf}%
\right)  $ would suffice. However, we leave the exact formulation to
demonstrate how the present approach does not require costly repricing.

Consider a portfolio represented by the vector $\mathbf{w}$, whose generic $i
$-th entry is the number of contracts in the respective butterfly. The p\&l
then reads%
\begin{equation}
\Pi_{\mathbf{w}}\equiv\sum_{i=1}^{I}w_{i}\left(  P_{i}\left(  \mathbf{X}%
,\mathcal{I}_{t}\right)  -P_{i,t}\right)  \text{,}%
\end{equation}
where $P_{i}\left(  \mathbf{X},\mathcal{I}_{t}\right)  $ is the price at the
horizon $\left(  \ref{PRasf}\right)  $ and $P_{i,t}$ is the currently traded
price of the $i$-th butterfly. We assume that, in order to account for market
asymmetries and downside risk, the trader optimizes the mean-CVaR trade-off.
Therefore $\left(  \ref{asgdsedhid}\right)  $ becomes%
\begin{equation}
\mathbf{w}_{\lambda}\equiv\operatorname*{argmax}_{\underline{\mathbf{b}}%
\leq\mathbf{Bw}\leq\overline{\mathbf{b}}}\left\{  \mathbb{E}\left\{
\Pi_{\mathbf{w}}\right\}  -\lambda\operatorname*{CVaR}\nolimits_{\gamma
}\left\{  \Pi_{\mathbf{w}}\right\}  \right\}
\text{,\label{asgdsedhid copy(5)}}%
\end{equation}
where $\gamma$ is the CVaR tail level; and $\mathbf{B}$, $\underline
{\mathbf{b}}$, and $\overline{\mathbf{b}}$ are a matrix and vectors that
represent investment constraints.

To illustrate, we set $\gamma\equiv95\%$, we impose that the long-short
positions offset to a zero delta and a zero initial budget, and that the
absolute investment in each option does not exceed a fixed threshold. We set
the investment horizon as $\tau\equiv1$ day. We consider a limited market of
$I\equiv9$ securities: 1-month, 2-month and 6-month butterflies on the three
technology stocks Microsoft (M), Yahoo (Y) and Google (G).

In addition to the respective underlyings and implied volatilities, we include
the possibility of views on growth or inflation, as represented by the slope
of the interest rate curve:\ therefore we add the changes in the two- and
ten-year points of the curve, for a total of $N\equiv14$ factors:%
\begin{equation}
\mathbf{X}\equiv\left(  X^{M},X_{1m}^{M},X_{2m}^{M},X_{6m}^{M},\ldots
,,X_{6m}^{G},X_{2y},X_{10y}\right)  ^{\prime}\text{.}%
\end{equation}
To determine the reference distribution $\left(  \ref{MAara}\right)  $ of
these factors we consider the panel of joint observations of the factors over
a three-year horizon: this amounts to $700$ observations. To achieve
$J\equiv10^{5}$ joint simulations we kernel-bootstrap the historical
scenarios:\ for each historical observation $\mathbf{x}_{t}$, we draw
$10^{5}/700$ observations from the multivariate normal distribution
$\operatorname{N}\left(  \mathbf{x}_{t},\epsilon\widehat{\mathbf{\Sigma}%
}\right)  $, where $\widehat{\mathbf{\Sigma}}$ is the sample covariance and we
set $\epsilon\equiv0.15$. The juxtaposition of the above simulations yields
the desired $J\times N$ panel $\mathcal{X}$, where each scenario has equal
probability $p_{j}\equiv1/J$.

Then we input each scenario of $\mathcal{X}$ into the pricing function
$\left(  \ref{PRicsnafg}\right)  $, obtaining the joint p\&l scenarios
$\mathcal{P}$ with equal probabilities $\mathbf{p}$. The sample counterpart of
the mean-CVaR efficient frontier $\left(  \ref{asgdsedhid copy(5)}\right)  $
reads%
\begin{equation}
\mathbf{w}_{\lambda}\equiv\operatorname*{argmax}_{\underline{\mathbf{b}}%
\leq\mathbf{Bw}\leq\overline{\mathbf{b}}}\left\{  \left(  \mathbf{w}^{\prime
}\mathcal{P}^{\prime}\mathbf{p}\right)  +\lambda\frac{\left[  \mathbf{p}%
\right]  ^{\prime}\left[  \mathcal{P}\mathbf{w}\right]  }{\left[
\mathbf{p}\right]  ^{\prime}\left[  \mathbf{1}\right]  }\right\}
\text{,\label{Inasgasdg copy(1)}}%
\end{equation}
where the operator $\left[  \mathbf{x}\right]  $ selects in the generic vector
$\mathbf{x}$ only the entries that correspond to the $\left(  1-\gamma\right)
J$ smallest entries of $\mathcal{P}\mathbf{w}$. If $J$ is not too large this
can be solved by linear programming as in \cite{RockafellarUryas00}. For very
large $J$ we solve this heuristically as in \cite{Meucci05} by a two-step
approach:\ first determine the mean-variance efficient frontier, then perform
a uni-variate grid search for the optimal trade-off $\left(
\ref{Inasgasdg copy(1)}\right)  $.%

\begin{figure}[h]%
\centering
\includegraphics[
height=141pt,
width=302.25pt
]%
{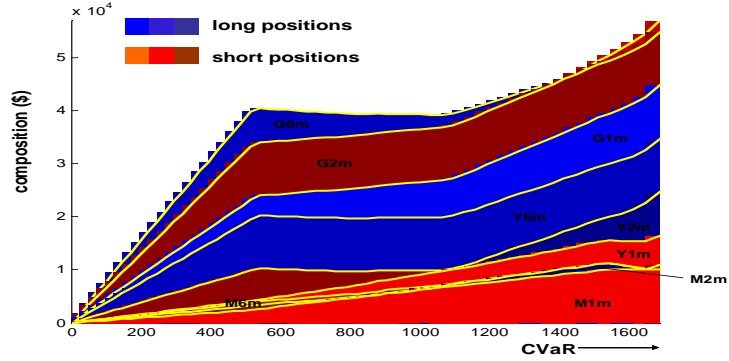}%
\caption{Mean-CVaR long-short efficient frontier: prior risk model}%
\label{FigPERiaf}%
\end{figure}
In Figure \ref{FigPERiaf} we display the frontier ensuing from the reference
market model in our example. For the extreme case of zero risk appetite, not
investing at all is optimal. As the risk appetite increases, leverage
increases, always respecting the constraint of a zero net initial investment,
as well as delta-neutrality. When the risk appetite increases further, the
remaining constraints enter the picture.

\begin{figure}[h]%
\centering
\includegraphics[
height=141pt,
width=302.25pt
]%
{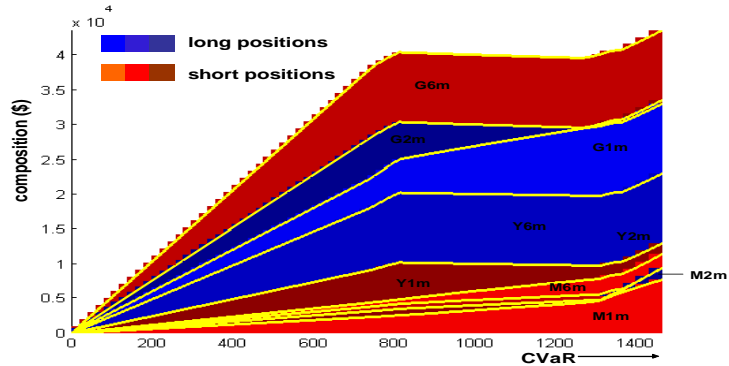}%
\caption{Mean-CVaR long-short efficient frontier:\ view on G6m-G2m spread}%
\label{FIgPOsateSPera}%
\end{figure}

Now we consider the views of three distinct analysts. The first one is bearish
about the 2m-6m implied volatility spread for Google. From $\left(
\ref{BUllwivas}\right)  $-$\left(  \ref{PSfadgdg}\right)  $ this means%
\begin{equation}
\widetilde{\mathbb{E}}\left\{  X_{6m}^{G}-X_{2m}^{G}\right\}  \leq
\mathbb{E}\left\{  X_{6m}^{G}-X_{2m}^{G}\right\}  -\sigma\left\{  X_{6m}%
^{G}-X_{2m}^{G}\right\}  \text{.\label{Vadidaada}}%
\end{equation}
This view is represented in the form $\left(  \ref{Cliagad}\right)  $ as%
\begin{equation}
\sum_{j=1}^{J}\widetilde{p}_{j}^{\left(  1\right)  }\left(  X_{j,6m}%
^{G}-X_{j,2m}^{G}\right)  \leq\widehat{m}_{6|2}-\widehat{\sigma}_{6|2}\text{,}
\label{Serasg}%
\end{equation}
where $\widehat{m}_{6|2}$ and $\widehat{\sigma}_{6|2}$ are the sample
counterparts of the respective terms in $\left(  \ref{Vadidaada}\right)  $. We
can compute $\widetilde{\mathbf{p}}^{\left(  1\right)  }$ as in $\left(
\ref{acsggad}\right)  $, under the constraint $\left(  \ref{Serasg}\right)  $.
To illustrate, we show In Figure \ref{FIgPOsateSPera} the mean-CVaR efficient
frontier $\left(  \ref{Inasgasdg copy(1)}\right)  $ when this view is
processed: as expected, the G6m-G2m spread, previously long, is now short.%

The second analyst is bullish on the realized volatility of Microsoft, defined
as $\left\vert X^{M}\right\vert $, the absolute log-change in the underlying:
this is the variable such that, if larger than a threshold, a long position in
the butterfly turns into a profit. Since this variable displays thick tails
and the expectation might not be defined, see e.g. \cite{rachev03}, we issue a
relative statement on the median, comparing it with the third quintile implied
by the reference market model:%
\begin{equation}
\widetilde{\mathbb{M}}\left\{  \left\vert X^{M}\right\vert \right\}  \geq
Q_{\left\vert X^{M}\right\vert }\left(  \frac{3}{5}\right)  \text{.}%
\end{equation}
This view is represented in the form $\left(  \ref{Cliagad}\right)  $ as%
\begin{equation}
\sum_{j\in\widetilde{J}}\widetilde{p}_{j}^{\left(  2\right)  }\leq\frac{1}%
{2}\text{,\label{VIdaeSsytw}}%
\end{equation}
where $\widetilde{J}$ is the set of indices $j$ such that $\left\vert
X_{j}^{M}\right\vert $ is smaller than the sample third quintile of
$\left\vert X^{M}\right\vert $, see Appendix \ref{APpdgiasdpg}. Now we can
compute $\widetilde{\mathbf{p}}^{\left(  2\right)  }$ as in $\left(
\ref{acsggad}\right)  $ under the constraint $\left(  \ref{VIdaeSsytw}\right)
$.

The third analyst believes that the slope of the curve will increase by five
basis points. Therefore he formulates the view a-la BL, using in $\left(
\ref{BUllwivas}\right)  $ expectations and binding constraints:%
\begin{equation}
\sum_{j=1}^{J}\widetilde{p}_{j}^{\left(  3\right)  }\left(  X_{j,10y}%
-X_{j,2y}\right)  \equiv0.0005\text{.}%
\end{equation}
and $\widetilde{\mathbf{p}}^{\left(  3\right)  }$ can be computed as in
$\left(  \ref{acsggad}\right)  $.%

\begin{figure}[h]%
\centering
\includegraphics[
height=141pt,
width=302.25pt
]%
{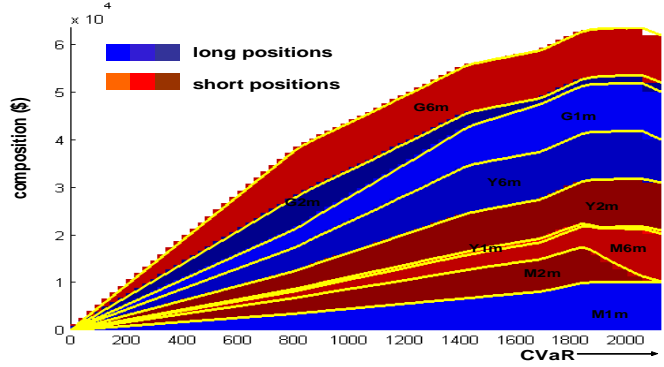}%
\caption{Mean-CVaR long-short efficient frontier:\ all views}%
\label{FIgPOsteStaea}%
\end{figure}
The management committee attributes $c_{1}\equiv0.20$, $c_{2}\equiv0.25$ and
$c_{3}\equiv0.20$ confidence on the analysts' views, the remaining portion
being attributed to the reference model. Then the uncertainty-weighted
posterior probabilities read
\begin{equation}
\widetilde{\mathbf{p}}_{\mathbf{c}}\equiv\sum_{s=0}^{3}c_{s}\widetilde
{\mathbf{p}}^{\left(  s\right)  }\text{,}%
\end{equation}
where $c_{0}\equiv1-c_{1}-c_{2}-c_{3}$ and $\widetilde{\mathbf{p}}^{\left(
0\right)  }\equiv\mathbf{p}$. We show in Figure \ref{FIgPOsteStaea} the
combined effects of all the views on the frontier $\left(
\ref{Inasgasdg copy(1)}\right)  $.

We emphasize that in this case study the market has a non-parametric,
thick-tailed, non-normal distribution; two views are expressed as
inequalities; one view acts on a non-linear function, the absolute value, of a
factor; the slope of the curve in one view is an external factor that appears
nowhere in the pricing function of the securities; features different from
expectations are being assessed, namely the median; and no repricing was ever necessary.

\section{Conclusions\label{SeccdCndogd}}

We present the EP, a unified framework to perform trading, portfolio
management and generalized stress-testing in markets with complex derivatives
driven by non-normal factors. The inputs are a possibly non-normal reference
market model and a set of very general equality or inequality views on a
variety of features of the market. The output is a posterior distribution that
incorporates all the inputs. As it turns out, the EP avoids costly repricing
by representing the posterior distribution in terms of the same scenarios as
the reference model, but with different probabilities whose computation is
extremely efficient.

We summarize in the table below the capabilities of the EP as compared to
\cite{BlackLitt90}, \cite{AlmgrenChriss06}, \cite{Qian01}, \cite{Pezier07},
\cite{Meucci08f} and the COP in \cite{Meucci06b}.%

\[%
\begin{tabular}
[c]{lccccccc}
& BL & AC & QG & P & M & COP & EP\\
normal market \& linear views & $\checkmark$ & $\cdot$ & $\checkmark$ &
$\checkmark$ & $\checkmark$ & $\checkmark$ & $\checkmark$\\
scenario analysis & $\cdot$ & $\cdot$ & $\checkmark$ & $\checkmark$ &
$\checkmark$ & $\checkmark$ & $\checkmark$\\
correlation stress-test & $\cdot$ & $\cdot$ & $\checkmark$ & $\checkmark$ &
$\checkmark$ & $\cdot$ & $\checkmark$\\
trading desk: non-linear pricing & $\cdot$ & $\cdot$ & $\cdot$ & $\cdot$ &
$\checkmark$ & $\checkmark$ & $\checkmark$\\
external factors: macro, etc. & $\cdot$ & $\cdot$ & $\cdot$ & $\cdot$ &
$\checkmark$ & $\checkmark$ & $\checkmark$\\
partial specifications & $\cdot$ & $\cdot$ & $\cdot$ & $\checkmark$ & $\cdot$
& $\cdot$ & $\checkmark$\\
non-normal market & $\cdot$ & $\cdot$ & $\cdot$ & $\cdot$ & $\cdot$ &
$\checkmark$ & $\checkmark$\\
multiple users & $\cdot$ & $\cdot$ & $\cdot$ & $\cdot$ & $\cdot$ &
$\checkmark$ & $\checkmark$\\
non-linear views & $\cdot$ & $\cdot$ & $\cdot$ & $\cdot$ & $\cdot$ & $\cdot$ &
$\checkmark$\\
trading desk: costly pricing & $\cdot$ & $\cdot$ & $\cdot$ & $\cdot$ & $\cdot$
& $\cdot$ & $\checkmark$\\
lax constraints: ranking & $\cdot$ & $\checkmark$ & $\cdot$ & $\cdot$ &
$\cdot$ & $\cdot$ & $\checkmark$%
\end{tabular}
\]

\bibliographystyle{jf}
\bibliography{Finance}

\newpage

\appendix

\section{Appendix}

In this appendix we present proofs, results and details that can be skipped at
first reading.

\subsection{The analytical solution\label{Appfasfppdfd}}

Using the explicit expression for the multivariate normal pdf
\begin{equation}
\ln f_{\mathbf{\mu},\mathbf{\Sigma}}\left(  \mathbf{x}\right)  \equiv-\frac
{N}{2}\ln\left(  2\pi\right)  -\frac{1}{2}\ln\left\vert \mathbf{\Sigma
}\right\vert -\frac{1}{2}\left(  \mathbf{x}-\mathbf{\mu}\right)  ^{\prime
}\Sigma^{-1}\left(  \mathbf{x}-\mathbf{\mu}\right)
\end{equation}
we can compute the Kullback-Leibler divergence between normal distributions:%
\begin{align}
D_{KL}\left(  f_{\widetilde{\mathbf{\mu}},\widetilde{\mathbf{\Sigma}}%
},f_{\mathbf{\mu},\mathbf{\Sigma}}\right)   &  \equiv\int_{\mathbb{R}^{N}%
}f_{\widetilde{\mathbf{\mu}},\widetilde{\mathbf{\Sigma}}}\left(
\mathbf{x}\right)  \ln f_{\widetilde{\mathbf{\mu}},\widetilde{\mathbf{\Sigma}%
}}\left(  \mathbf{x}\right)  d\mathbf{x}\text{\label{KSagfs}}\\
&  -\int_{\mathbb{R}^{N}}f_{\widetilde{\mathbf{\mu}},\widetilde{\mathbf{\Sigma
}}}\left(  \mathbf{x}\right)  \ln f_{\mathbf{\mu},\mathbf{\Sigma}}\left(
\mathbf{x}\right)  d\mathbf{x}\nonumber\\
&  =-\frac{N}{2}\ln\left(  2\pi\right)  -\frac{1}{2}\ln\left\vert
\widetilde{\mathbf{\Sigma}}\right\vert -\frac{1}{2}\widetilde{\mathbb{E}%
}\left\{  \left(  \mathbf{X}-\widetilde{\mathbf{\mu}}\right)  ^{\prime
}\widetilde{\mathbf{\Sigma}}^{-1}\left(  \mathbf{X}-\widetilde{\mathbf{\mu}%
}\right)  \right\} \nonumber\\
&  +\frac{N}{2}\ln\left(  2\pi\right)  +\frac{1}{2}\ln\left\vert
\mathbf{\Sigma}\right\vert +\frac{1}{2}\widetilde{\mathbb{E}}\left\{  \left(
\mathbf{X}-\mathbf{\mu}\right)  ^{\prime}\mathbf{\Sigma}^{-1}\left(
\mathbf{X}-\mathbf{\mu}\right)  \right\} \nonumber\\
&  =\frac{1}{2}\ln\left\vert \widetilde{\mathbf{\Sigma}}^{-1}\mathbf{\Sigma
}\right\vert -\frac{1}{2}\operatorname{tr}\left[  \widetilde{\mathbb{E}%
}\left\{  \left(  \mathbf{X}-\widetilde{\mathbf{\mu}}\right)  \left(
\mathbf{X}-\widetilde{\mathbf{\mu}}\right)  ^{\prime}\right\}  \widetilde
{\mathbf{\Sigma}}^{-1}\right] \nonumber\\
&  +\frac{1}{2}\operatorname{tr}\left[  \widetilde{\mathbb{E}}\left\{  \left(
\mathbf{X}-\mathbf{\mu}\right)  \left(  \mathbf{X}-\mathbf{\mu}\right)
^{\prime}\right\}  \mathbf{\Sigma}^{-1}\right] \nonumber\\
&  =\frac{1}{2}\ln\left\vert \widetilde{\mathbf{\Sigma}}^{-1}\mathbf{\Sigma
}\right\vert -\frac{N}{2}+\frac{1}{2}\operatorname{tr}\left[  \left(
\widetilde{\mathbf{\Sigma}}+\left(  \widetilde{\mathbf{\mu}}-\mathbf{\mu
}\right)  \left(  \widetilde{\mathbf{\mu}}-\mathbf{\mu}\right)  ^{\prime
}\right)  \mathbf{\Sigma}^{-1}\right] \nonumber\\
&  =\frac{1}{2}\ln\left\vert \widetilde{\mathbf{\Sigma}}^{-1}\mathbf{\Sigma
}\right\vert -\frac{N}{2}+\frac{1}{2}\operatorname{tr}\left[  \widetilde
{\mathbf{\Sigma}}\mathbf{\Sigma}^{-1}\right] \nonumber\\
&  +\frac{1}{2}\left(  \widetilde{\mathbf{\mu}}-\mathbf{\mu}\right)  ^{\prime
}\mathbf{\Sigma}^{-1}\left(  \widetilde{\mathbf{\mu}}-\mathbf{\mu}\right)
\nonumber
\end{align}

Our purpose is to minimize the Kullback-Leibler divergence $\left(
\ref{KSagfs}\right)  $ under the constraints $\left(  \ref{Coaidsgna}\right)
$. Using the following matrix identity%
\begin{equation}
\operatorname{vec}\left(  \mathbf{\Gamma}\right)  ^{\prime}\operatorname{vec}%
\left(  \mathbf{A}\right)  \equiv\sum_{i,k}\Gamma_{ki}A_{ki}=\operatorname{tr}%
\left(  \mathbf{\Gamma}^{\prime}\mathbf{A}\right)  \text{,\label{casiofadso}}%
\end{equation}
we write the Lagrangian as%
\begin{align}
\mathcal{L}  &  =\frac{1}{2}\left(  \widetilde{\mathbf{\mu}}-\mathbf{\mu
}\right)  ^{\prime}\mathbf{\Sigma}^{-1}\left(  \widetilde{\mathbf{\mu}%
}-\mathbf{\mu}\right)  +\frac{1}{2}\operatorname{tr}\left(  \mathbf{\Sigma
}^{-1}\widetilde{\mathbf{\Sigma}}\right)  -\frac{1}{2}\ln\left(  \left\vert
\mathbf{\Sigma}^{-1}\widetilde{\mathbf{\Sigma}}\right\vert \right) \\
&  -\mathbf{\lambda}^{\prime}\left(  \mathbf{Q}\widetilde{\mathbf{\mu}%
}-\widetilde{\mathbf{\mu}}_{\mathbf{Q}}\right)  -\frac{1}{2}\operatorname{tr}%
\left(  \mathbf{\Gamma}^{\prime}\left(  \mathbf{G}\widetilde{\mathbf{\Sigma}%
}\mathbf{G}^{\prime}-\widetilde{\mathbf{\Sigma}}\mathbf{_{\mathbf{G}}}\right)
\right)  \text{.}\nonumber
\end{align}

The first order conditions for $\widetilde{\mathbf{\mu}}$ read%
\begin{equation}
\mathbf{0}\equiv\frac{\partial\mathcal{L}}{\partial\widetilde{\mathbf{\mu}}%
}=\mathbf{\Sigma}^{-1}\left(  \widetilde{\mathbf{\mu}}-\mathbf{\mu}\right)
-\mathbf{Q}^{\prime}\mathbf{\lambda}\text{,}%
\end{equation}
or equivalently%
\begin{equation}
\widetilde{\mathbf{\mu}}-\mathbf{\mu}=\mathbf{\Sigma Q}^{\prime}%
\mathbf{\lambda}\text{.\label{Faidogg}}%
\end{equation}
Pre-multiplying by $\mathbf{Q}$ both sides this implies%
\begin{equation}
\mathbf{\lambda}=\left(  \mathbf{Q\Sigma Q}^{\prime}\right)  ^{-1}\left(
\widetilde{\mathbf{\mu}}_{\mathbf{Q}}-\mathbf{Q\mu}\right)  \text{.}%
\end{equation}
Substituting this in $\left(  \ref{Faidogg}\right)  $ we obtain%
\begin{equation}
\widetilde{\mathbf{\mu}}=\mathbf{\mu+\Sigma Q}^{\prime}\left(  \mathbf{Q}%
\widetilde{\mathbf{\Sigma}}\mathbf{Q}^{\prime}\right)  ^{-1}\left(
\widetilde{\mathbf{\mu}}_{\mathbf{Q}}-\mathbf{Q\mu}\right)  \label{gfpaogjad}%
\end{equation}

To determine the first order conditions for $\widetilde{\mathbf{\Sigma}}$ we
first use the identity in \cite{Minka03}%
\begin{equation}
d\ln\left\vert \mathbf{X}\right\vert =\operatorname{tr}\left(  \mathbf{X}%
^{-1}d\mathbf{X}\right)
\end{equation}
and the symmetry of $\mathbf{\Gamma}$ to express the differential of the
Lagrangian with respect to $\widetilde{\mathbf{\Sigma}}$ as follows:%
\begin{equation}
d\mathcal{L}=\frac{1}{2}\operatorname{tr}\left(  \mathbf{\Sigma}%
^{-1}d\widetilde{\mathbf{\Sigma}}\right)  -\frac{1}{2}\operatorname{tr}\left(
\widetilde{\mathbf{\Sigma}}^{-1}d\widetilde{\mathbf{\Sigma}}\right)  -\frac
{1}{2}\operatorname{tr}\left(  \mathbf{G}^{\prime}\mathbf{\Gamma G}%
d\widetilde{\mathbf{\Sigma}}\right)  \text{.\label{asjgpasdg}}%
\end{equation}
Using again $\left(  \ref{casiofadso}\right)  $ to setting $\left(
\ref{asjgpasdg}\right)  $ to zero we obtain:%
\begin{equation}
\widetilde{\mathbf{\Sigma}}^{-1}=\mathbf{\Sigma}^{-1}-\mathbf{G}^{\prime
}\mathbf{\Gamma G}\text{\label{aospgjpdag}}%
\end{equation}
Using the following matrix identity ($\mathbf{A}$ and $\mathbf{D}$ invertible,
$\mathbf{B}$ and $\mathbf{C}$ conformable)%
\begin{equation}
\left(  \mathbf{A}-\mathbf{BD}^{-1}\mathbf{C}\right)  ^{-1}=\mathbf{A}%
^{-1}-\mathbf{A}^{-1}\mathbf{B}\left(  \mathbf{CA}^{-1}\mathbf{B}%
-\mathbf{D}\right)  ^{-1}\mathbf{CA}^{-1}\text{,} \label{MatrIdent87}%
\end{equation}
we can write $\left(  \ref{aospgjpdag}\right)  $ as
\begin{align}
\widetilde{\mathbf{\Sigma}}  &  =\left(  \mathbf{\Sigma}^{-1}-\mathbf{G}%
^{\prime}\mathbf{\Gamma G}\right)  ^{-1}\label{apogjpadgd}\\
&  =\mathbf{\Sigma}-\mathbf{\Sigma G}^{\prime}\left(  \mathbf{G\Sigma
G}^{\prime}-\mathbf{\Gamma}^{-1}\right)  ^{-1}\mathbf{G\Sigma}\text{.}%
\nonumber
\end{align}
Using the constraints%
\begin{equation}
\widetilde{\mathbf{\Sigma}}\mathbf{_{\mathbf{G}}}\equiv\mathbf{G}%
\widetilde{\mathbf{\Sigma}}\mathbf{G}^{\prime}=\mathbf{G\Sigma G}^{\prime
}-\mathbf{G\Sigma G}^{\prime}\left(  \mathbf{G\Sigma G}^{\prime}%
-\mathbf{\Gamma}^{-1}\right)  ^{-1}\mathbf{G\Sigma G}^{\prime}%
\end{equation}
or%
\begin{equation}
\left(  \mathbf{G\Sigma G}^{\prime}-\mathbf{\Gamma}^{-1}\right)  ^{-1}=\left(
\mathbf{G\Sigma G}^{\prime}\right)  ^{-1}-\left(  \mathbf{G\Sigma G}^{\prime
}\right)  ^{-1}\widetilde{\mathbf{\Sigma}}\mathbf{_{\mathbf{G}}}\left(
\mathbf{G\Sigma G}^{\prime}\right)  ^{-1}%
\end{equation}
Substituting this result back into $\left(  \ref{apogjpadgd}\right)  $ yields%
\begin{equation}
\widetilde{\mathbf{\Sigma}}=\mathbf{\Sigma}+\mathbf{\Sigma G}^{\prime}\left(
\left(  \mathbf{G\Sigma G}^{\prime}\right)  ^{-1}\widetilde{\mathbf{\Sigma}%
}\mathbf{_{\mathbf{G}}}\left(  \mathbf{G\Sigma G}^{\prime}\right)
^{-1}-\left(  \mathbf{G\Sigma G}^{\prime}\right)  ^{-1}\right)
\mathbf{G\Sigma}\text{.}%
\end{equation}

\subsection{Views as linear constraints on the
probabilities\label{APpdgiasdpg}}

Since this change is fully defined by the reference and the posterior
distribution of the views $\mathbf{V}$, to determine $\widetilde{\mathbf{p}}$
we need only focus on this lower dimensional space instead of the whole market
$\mathbf{X}$.

\subsubsection{Partial information views}

\begin{itemize}
\item Views a-la Black Litterman
\end{itemize}

The generalized BL bullish/bearish view reads%
\begin{equation}
\widetilde{m}\left\{  V_{k}\right\}  \gtreqqless m_{k}%
\text{.\label{BUllwivasss}}%
\end{equation}

We can define $m_{k}$ exogenously. Alternatively, as in $\left(
\ref{PSfadgdg}\right)  $ we set
\begin{equation}
m_{k}\equiv\widehat{m}_{k}+\varkappa\widehat{\sigma}_{k}\text{,}%
\end{equation}
where $\widehat{m}_{k}$ is the sample mean of the $k$-th column of the panel
$\mathcal{V}$ based on the prior probability
\begin{equation}
\widehat{m}_{k}\equiv\sum_{j=1}^{J}p_{j}\mathcal{V}_{j,k}\text{,}%
\end{equation}
and $\widehat{\sigma}_{k}$ is its sample standard deviation of the $k$-th
column of the panel $\mathcal{V}$ based on the prior probability
\begin{equation}
\widehat{\sigma}_{k}^{2}\equiv\sum_{j=1}^{J}p_{j}\left(  \mathcal{V}%
_{j,k}-\widehat{m}_{k}\right)  ^{2}\text{.}%
\end{equation}

Alternatively, we set $m_{k}$ in $\left(  \ref{BUllwivasss}\right)  $ as the
sample $\left(  \frac{1}{2}+\frac{\kappa}{5}\right)  $-tile of the $k$-th
column of the panel $\mathcal{V}$ based on the prior probability%
\begin{equation}
m_{k}\equiv\mathcal{V}_{s\left(  \overline{I}\right)  ,k}%
\text{.\label{Sadgpog}}%
\end{equation}
In this expression $s$ is the sorting function of the $k$-th column of the
panel $\mathcal{V}$, i.e. denoting by $\mathcal{V}_{i:J,k}$ the $i$-th order
statistics of the $k$-th column the function $s$ is defined as%
\begin{equation}
\mathcal{V}_{s\left(  i\right)  ,k}\equiv\mathcal{V}_{i:J,k}\text{,\quad
}i=1,\ldots,J\text{;}%
\end{equation}
and the index $\overline{I}$ satisfies%
\begin{equation}
\overline{I}\equiv\operatorname*{argmax}_{I}\left\{  \sum_{i=1}^{I}p_{s\left(
i\right)  }\leq\left(  \frac{1}{2}+\frac{\kappa}{5}\right)  \right\}  \text{.}%
\end{equation}

To express $\left(  \ref{BUllwivasss}\right)  $ as in $\left(  \ref{Cliagad}%
\right)  $ we first consider the case where $\widetilde{m}\left\{
V_{k}\right\}  $ is the expectation. Then its sample counterpart is the sample
mean and $\left(  \ref{BUllwivasss}\right)  $ reads%
\begin{equation}
\sum_{j=1}^{J}\widetilde{p}_{j}\mathcal{V}_{j,k}\gtreqqless m_{k}\text{,}%
\end{equation}

On the other hand, if $\widetilde{m}\left\{  V_{k}\right\}  $ in $\left(
\ref{BUllwivasss}\right)  $ is the median, then the view reads%
\begin{equation}
\sum_{j\in I_{k}}\widetilde{p}_{j}\gtreqqless\frac{1}{2}\text{,}%
\end{equation}
where $I_{k}$ denotes the indices of the scenarios in $\mathcal{V}_{\cdot,k} $
larger than $m_{k}$.

\begin{itemize}
\item Relative ranking
\end{itemize}

The relative ordering view%
\begin{equation}
\widetilde{m}\left\{  V_{1}\right\}  \geq\widetilde{m}\left\{  V_{2}\right\}
\geq\cdots\geq\widetilde{m}\left\{  V_{K}\right\}  \text{,}
\label{SyoafkaRand copy(1)}%
\end{equation}
when the location parameter is expectation translates into the following set
of linear constraints:%
\begin{align}
\sum_{j=1}^{J}\widetilde{p}_{j}\left(  \mathcal{V}_{j,1}-\mathcal{V}%
_{j,2}\right)   &  \geq0\nonumber\\
&  \vdots\\
\sum_{j=1}^{J}\widetilde{p}_{j}\left(  \mathcal{V}_{j,K-1}-\mathcal{V}%
_{j,K}\right)   &  \geq0\text{.}\nonumber
\end{align}

\begin{itemize}
\item Views on volatility
\end{itemize}

A view on volatility reads%
\begin{equation}
\widetilde{\sigma}\left\{  V_{k}\right\}  \gtreqqless\sigma_{k}%
\text{.\label{tasbVOslafs copy(1)}}%
\end{equation}

First we consider the case where $\widetilde{\sigma}\left\{  V_{k}\right\}  $
is the standard deviation. Then $\left(  \ref{tasbVOslafs copy(1)}\right)  $
can be expressed as in $\left(  \ref{Cliagad}\right)  $ as%
\begin{equation}
\sum_{j=1}^{J}\widetilde{p}_{j}\mathcal{V}_{j,k}^{2}\gtreqqless\widehat{m}%
_{k}^{2}+\sigma_{k}^{2}\text{,}%
\end{equation}
where $\widehat{m}_{k}$ is the sample mean of the $k$-th column of the panel
$\mathcal{V}$. The benchmark $\sigma_{k}$ can be set exogenously.
Alternatively, we set
\begin{equation}
\sigma_{k}\equiv\varkappa\widehat{\sigma}_{k}\text{,}%
\end{equation}
where $\widehat{\sigma}_{k}$ is the sample standard deviation of the $k$-th
column of the panel $\mathcal{V}$.

When $\widetilde{\sigma}\left\{  V_{k}\right\}  $ in $\left(
\ref{tasbVOslafs copy(1)}\right)  $ is the range between the $\left(  \frac
{1}{2}-\gamma\right)  $-tile and the $\left(  \frac{1}{2}+\gamma\right)
$-tile of the distribution of $V_{k}$ we proceed as follows. First, compute
the sample $\left(  \frac{1}{2}-\kappa\gamma\right)  $-tile $\underline
{\mathcal{V}}_{k}$ of the $k$-th column of the panel $\mathcal{V}$ as in
$\left(  \ref{Sadgpog}\right)  $ and similarly the sample $\left(  \frac{1}%
{2}+\kappa\gamma\right)  $-tile $\overline{\mathcal{V}}_{k}$. Then the view
reads%
\begin{equation}
\sum_{j\in\underline{I}_{k}}\widetilde{p}_{j}\gtreqqless\frac{1}{2}%
-\gamma,\quad\sum_{j\in\overline{I}_{k}}\widetilde{p}_{j}\gtreqqless\frac
{1}{2}-\gamma\text{.}%
\end{equation}
where $\underline{I}_{k}$ denotes the scenarios in the $k$-th column of
$\mathcal{V}$ that are smaller than $\underline{\mathcal{V}}_{k}$ and
$\underline{I}_{k}$ denotes the scenarios that are larger than $\overline
{\mathcal{V}}_{k}$.

\begin{itemize}
\item Views on correlations
\end{itemize}

To stress test the correlations with a pre-defined matrix such as $\left(
\ref{COtdefs}\right)  $ we impose%
\begin{equation}
\sum_{j=1}^{J}\widetilde{p}_{j}\mathcal{V}_{j,k}\mathcal{V}_{j,l}%
\equiv\widehat{m}_{k}\widehat{m}_{l}+\widehat{\sigma}_{k}\widehat{\sigma}%
_{l}\widetilde{\mathbb{C}}_{k,l}\text{,}%
\end{equation}
where $\widehat{m}_{k}$ is the sample mean and $\widehat{\sigma}_{k}$ is the
sample standard deviation of the $k$-th column of the panel $\mathcal{V}$ .

\begin{itemize}
\item Views on tail codependence
\end{itemize}

First we extract the empirical copula from the panel $\mathcal{V}$ as in
\cite{Meucci06b}: we sort the columns of $\mathcal{V}$ in ascending order;
then we define a panel $\mathcal{U}$, whose generic $\left(  j,k\right)  $-th
entry is the normalized ranking of $\mathcal{V}_{j,k}$\ within the $k$-th
column (for instance, if $\mathcal{V}_{5,7}$ is the 423-th smallest simulation
in column $7$, then $\mathcal{U}_{5,7}\equiv423/J$). Each row of $\mathcal{U}$
represents a simulation from the copula of $f_{\mathbf{V}}$.

Stress-testing the tail codependence means%
\begin{equation}
\widetilde{C}_{\mathbf{V}}\left(  \mathbf{u}\right)  \gtreqqless\widetilde
{C}\text{,\label{GTIsalzidgd copy(2)}}%
\end{equation}
where $\widetilde{C}$ can be set exogenously. This translates into%
\begin{equation}
\sum_{j\in I_{\mathbf{u}}}\widetilde{p}_{j}\gtreqqless\widetilde{C}\text{,}%
\end{equation}
where $I_{\mathbf{u}}$ denotes the scenarios in $\mathcal{U}$ that lie jointly
below $\mathbf{u}$. To better tweak $\widetilde{C}$ a convenient formulation
is as the sample counterpart of $\varkappa C_{\mathbf{V}}\left(
\mathbf{u}\right)  $, for a reference copula $C_{\mathbf{V}}$ computed as above.

\subsubsection{Full-information views}

\begin{itemize}
\item Views on copula
\end{itemize}

If a full copula is specified, we draw a $J\times K$ panel of simulations
$\widetilde{\mathcal{U}}$ from it. To do so, we can fit to $\mathcal{U}$ a
parametric copula $\mathbf{U}_{\mathbf{\theta}}$ that depends on a set of
parameters $\mathbf{\theta}$; then $\widetilde{\mathcal{U}}$ is obtained by
drawing from the copula $\mathbf{U}_{\widetilde{\mathbf{\theta}}}$, where
$\widetilde{\mathbf{\theta}}$ is a perturbation of estimated parameters
$\mathbf{\theta}$.

Then $\widetilde{\mathbf{p}}$ is determined by matching all the cross moments
\begin{align}
\sum_{j=1}^{J}\widetilde{p}_{j}\mathcal{U}_{j,k}\mathcal{U}_{j,l}  &
=\sum_{j=1}^{J}p_{j}\widetilde{\mathcal{U}}_{j,k}\widetilde{\mathcal{U}}%
_{j,l},\quad k>l=1,\ldots,K\\
\sum_{j=1}^{J}\widetilde{p}_{j}\mathcal{U}_{j,k}\mathcal{U}_{j,l}%
\mathcal{U}_{j,i}  &  =\sum_{j=1}^{J}p_{j}\widetilde{\mathcal{U}}%
_{j,k}\widetilde{\mathcal{U}}_{j,l}\widetilde{\mathcal{U}}_{j,i},\quad
k>l>i=1,\ldots,K\\
&  \vdots\nonumber
\end{align}
and as well as all the marginal moments of the uniform distribution%
\begin{align}
\sum_{j=1}^{J}\widetilde{p}_{j}\mathcal{U}_{j,k}  &  =\frac{1}{2}\\
\sum_{j=1}^{J}\widetilde{p}_{j}\mathcal{U}_{j,k}^{2}  &  =\frac{1}{3}\\
&  \vdots\nonumber
\end{align}
up to a given order.

\begin{itemize}
\item Views on marginal distributions
\end{itemize}

If a full marginal distribution for the $k$-th view is specified, we draw a
$J\times1$ vector of simulations $\widetilde{\mathcal{V}}_{\cdot,k}$ from it.
Then $\widetilde{\mathbf{p}}$ is determined by matching all the moments up to
a given order:%
\begin{align}
\sum_{j=1}^{J}\widetilde{p}_{j}\mathcal{V}_{j,k}  &  =\sum_{j=1}^{J}%
p_{j}\widetilde{\mathcal{V}}_{j,k},\\
\sum_{j=1}^{J}\widetilde{p}_{j}\left(  \mathcal{V}_{j,k}\right)  ^{2}  &
=\sum_{j=1}^{J}p_{j}\left(  \widetilde{\mathcal{V}}_{j,k}\right)  ^{2}\\
\sum_{j=1}^{J}\widetilde{p}_{j}\left(  \mathcal{V}_{j,k}\right)  ^{3}  &
=\sum_{j=1}^{J}p_{j}\left(  \widetilde{\mathcal{V}}_{j,k}\right)  ^{3}\\
&  \vdots\nonumber
\end{align}

\begin{itemize}
\item Views on joint distribution
\end{itemize}

If a full joint view distribution $\left(  \ref{MAfa copy(1)}\right)  $ is
specified, we draw a $J\times K$ panel of simulations $\widetilde{\mathcal{V}%
}$ from it. This can be done in one shot, or by paring a desired copula with
desired marginals as in \cite{Meucci06b}. Then $\widetilde{\mathbf{p}}$ is
determined by matching all the cross moments up to a given order:%
\begin{align}
\sum_{j=1}^{J}\widetilde{p}_{j}\mathcal{V}_{j,k}  &  =\sum_{j=1}^{J}%
p_{j}\widetilde{\mathcal{V}}_{j,k},\quad k=1,\ldots,K\\
\sum_{j=1}^{J}\widetilde{p}_{j}\mathcal{V}_{j,k}\mathcal{V}_{j,l}  &
=\sum_{j=1}^{J}p_{j}\widetilde{\mathcal{V}}_{j,k}\widetilde{\mathcal{V}}%
_{j,l},\quad k\geq l=1,\ldots,K\\
\sum_{j=1}^{J}\widetilde{p}_{j}\mathcal{V}_{j,k}\mathcal{V}_{j,l}%
,\mathcal{V}_{j,i}  &  =\sum_{j=1}^{J}p_{j}\widetilde{\mathcal{V}}%
_{j,k}\widetilde{\mathcal{V}}_{j,l}\widetilde{\mathcal{V}}_{j,i},\quad k\geq
l\geq i=1,\ldots,K\\
&  \vdots\nonumber
\end{align}

\subsection{Numerical entropy minimization\label{AppSovEBter}}

The entropy minimization problem $\left(  \ref{acsggad}\right)  $ reads
explicitly%
\begin{equation}
\widetilde{\mathbf{p}}\equiv\operatorname*{argmin}_{\substack{\mathbf{Fx}%
\leq\mathbf{f} \\\mathbf{Hx}\equiv\mathbf{h}}}\left\{  \sum_{j=1}^{J}%
x_{j}\left(  \ln\left(  x_{j}\right)  -\ln\left(  p_{j}\right)  \right)
\right\}  \text{,\label{acsggad copy(1)}}%
\end{equation}
where we have collected all the inequality constraints in the matrix-vector
pair $\left(  \mathbf{F},\mathbf{f}\right)  $, all the equality constraints in
the matrix-vector pair $\left(  \mathbf{H},\mathbf{h}\right)  $ and where we
do not include the extra-constraint%
\begin{equation}
\mathbf{x}\geq\mathbf{0} \label{VCPsinsdbf}%
\end{equation}
because it will be automatically satisfied.

The Lagrangian for $\left(  \ref{acsggad copy(1)}\right)  $ reads%
\begin{equation}
\mathcal{L}\left(  \mathbf{x},\mathbf{\lambda},\mathbf{\nu}\right)
\equiv\mathbf{x}^{\prime}\left(  \ln\left(  \mathbf{x}\right)  -\ln\left(
\mathbf{p}\right)  \right)  +\mathbf{\lambda}^{\prime}\left(  \mathbf{Fx}%
-\mathbf{f}\right)  +\mathbf{\nu}^{\prime}\left(  \mathbf{Hx}-\mathbf{h}%
\right)  \text{.}%
\end{equation}
The first order conditions for $\mathbf{x}$ read%
\begin{equation}
\mathbf{0}\equiv\frac{\partial\mathcal{L}}{\partial\mathbf{x}}=\ln\left(
\mathbf{x}\right)  -\ln\left(  \mathbf{p}\right)  +\mathbf{1}+\mathbf{F}%
^{\prime}\mathbf{\lambda}+\mathbf{H}^{\prime}\mathbf{\nu}\text{.}%
\end{equation}
The solution is%
\begin{equation}
\mathbf{x}\left(  \mathbf{\lambda},\mathbf{\nu}\right)  =e^{\ln\left(
\mathbf{p}\right)  -\mathbf{1}-\mathbf{F}^{\prime}\mathbf{\lambda}%
-\mathbf{H}^{\prime}\mathbf{\nu}}\text{.}%
\end{equation}
Notice that the solution is always positive, which justifies not considering
$\left(  \ref{VCPsinsdbf}\right)  $.

The Lagrange dual function is defined as%
\begin{equation}
\mathcal{G}\left(  \mathbf{\lambda},\mathbf{\nu}\right)  \equiv\mathcal{L}%
\left(  \mathbf{x}\left(  \mathbf{\lambda},\mathbf{\nu}\right)
,\mathbf{\lambda},\mathbf{\nu}\right)  \text{.}%
\end{equation}
This function can be computed explicitly. The optimal Lagrange multipliers
follow from the numerical maximization of the Lagrange dual function%
\begin{equation}
\left(  \mathbf{\lambda}^{\ast},\mathbf{\nu}^{\ast}\right)  \equiv
\operatorname*{argmax}_{\mathbf{\lambda}\geq\mathbf{0},\mathbf{\nu}}\left\{
\mathcal{G}\left(  \mathbf{\lambda},\mathbf{\nu}\right)  \right\}  \text{.}
\label{NUmSOPgagds}%
\end{equation}
Notice that, whereas the Lagrangian should be minimized, the dual Lagrangian
must be maximized. Also notice that both gradient and Hessian can be easily
computed (the former from the envelope theorem) in order to speed up the
efficiency of the algorithm.

Finally, the solution to the original problem $\left(  \ref{acsggad copy(1)}%
\right)  $ reads%
\begin{equation}
\widetilde{\mathbf{p}}=\mathbf{x}\left(  \mathbf{\lambda}^{\ast},\mathbf{\nu
}^{\ast}\right)  \text{.}%
\end{equation}
The numerical optimization $\left(  \ref{NUmSOPgagds}\right)  $ acts on a very
limited number of variables, equal to the number of views. It does not act
directly on the very large number of variables of interest, namely the
probabilities of the Monte Carlo scenarios: this feature guarantees the
numerical feasibility of entropy optimization.

\subsection{Confidence specification\label{AppPRvab}}

We consider five increasingly complex cases. First, there is only one user
with equal confidence in all his views. Second, there is only one user, but
each view can potentially have a different confidence. Third, there are
multiple users, where each user has equal confidence in their own views.
Fourth, there are multiple users, but each view of each user can potentially
have a different confidence. Fifth, we propose a general framework to
accommodate all possible specifications.

\subsubsection{One user, equal confidence in all views}

This is the case considered in the pooling expression $\left(  \ref{Sateg}%
\right)  $. The confidence $c$ can be interpreted as the subjective
probability that the views be correct, instead of the reference market model.
Indeed, consider the mixture market%
\begin{equation}
\widehat{\mathbf{X}}\overset{d}{=}\left(  1-B\right)  \mathbf{X}%
+B\widetilde{\mathbf{X}}\text{,\label{PRoabvas}}%
\end{equation}
where $\mathbf{X}$ is distributed according to the reference model $\left(
\ref{MAara}\right)  $ and $\widetilde{\mathbf{X}}$ according to the regime
shift $\left(  \ref{FUlagsgas}\right)  $ implied by the views. If $B$ is a
$0$-$1$ Bernoulli variable that decides between the two regimes with
probabilities $1-c$ and $c$ respectively, the pdf of $\widehat{\mathbf{X}}$ is
exactly $\left(  \ref{Sateg}\right)  $.

Alternatively, we can represent the Bernoulli variable in $\left(
\ref{PRoabvas}\right)  $ as follows:%
\begin{equation}
\widehat{\mathbf{X}}\overset{d}{=}I_{1-c}\left(  U\right)  \mathbf{X}%
+I_{c}\left(  U\right)  \widetilde{\mathbf{X}}\text{,\label{PRoabvas copy(3)}}%
\end{equation}
where $U$ is a uniform random variable; and $I_{c}$ and $I_{1-c}$ are
indicator functions of non-overlapping intervals of size $c$ and $1-c$.

\subsubsection{One user, views with different confidences}

Consider the case where different views have different confidence levels. Each
view is a statement such as $\left(  \ref{BUllwivas}\right)  $-$\left(
\ref{GTIsalzidgd}\right)  $.

We illustrate this situation with an example%
\begin{equation}%
\begin{tabular}
[c]{ccc}%
index & view & confidence\\
$1$ & $\widetilde{m}\left\{  V_{1}\right\}  \geq\widetilde{m}\left\{
V_{2}\right\}  $ & $10\%$\\
$2$ & $\widetilde{m}\left\{  V_{2}\right\}  \geq\widetilde{m}\left\{
V_{3}\right\}  $ & $30\%$%
\end{tabular}
\end{equation}
One could model this situation in a way similar to $\left(  \ref{PRoabvas}%
\right)  $: in $10\%$ of the cases only the first view is satisfied and in
$30\%$ of the cases only the second view satisfied. However, this is not
correct. Instead, in $10\%$ of the cases both views are satisfied and in
$20\%$ of the cases only the second view is satisfied.

In other words, we are assigning probabilities to the subsets of views
combinations as follows:%
\begin{equation}%
\begin{tabular}
[c]{cc}%
subset & confidence\\
$\left\{  1,2\right\}  $ & $c_{\left\{  1,2\right\}  }\equiv10\%$\\
$\left\{  1\right\}  $ & $c_{\left\{  1\right\}  }\equiv0\%$\\
$\left\{  2\right\}  $ & $c_{\left\{  2\right\}  }\equiv20\%$\\
$\varnothing$ & $c_{\varnothing}\equiv70\%$%
\end{tabular}
\label{tqsabsaas}%
\end{equation}
Then, the posterior reads%
\begin{equation}
\widetilde{\mathbf{X}}\overset{d}{=}I_{c_{\varnothing}}\left(  U\right)
\widetilde{\mathbf{X}}_{\varnothing}+I_{c_{\left\{  1\right\}  }}\left(
U\right)  \widetilde{\mathbf{X}}_{\left\{  1\right\}  }+I_{c_{\left\{
2\right\}  }}\left(  U\right)  \widetilde{\mathbf{X}}_{\left\{  2\right\}
}+I_{c_{\left\{  1,2\right\}  }}\left(  U\right)  \widetilde{\mathbf{X}%
}_{\left\{  1,2\right\}  }\text{.\label{PRoabvas copy(1)}}%
\end{equation}
In this expression $\widetilde{\mathbf{X}}_{\varnothing}$ is a random variable
distributed according to the reference model $\left(  \ref{MAara}\right)  $;
$\widetilde{\mathbf{X}}_{\left\{  1\right\}  }$ is an independent random
variable, distributed according to the posterior with only the first view,
whose pdf, which follows from $\left(  \ref{FLSKnadd copy(1)}\right)  $, we
denote by $\widetilde{f}_{\left\{  1\right\}  }$; similarly for $\widetilde
{\mathbf{X}}_{\left\{  2\right\}  }$; $\widetilde{\mathbf{X}}_{\left\{
1,2\right\}  }$ is an independent random variable, distributed according to
the posterior from both views, whose pdf we denote by $\widetilde{f}_{\left\{
1,2\right\}  }$; $U$ is a uniform random variable; and the $I_{c}$'s are
indicators functions of the on non-overlapping intervals with size $c$ as in
Table \ref{tqsabsaas}: in particular $I_{c_{\left\{  1\right\}  }}\left(
U\right)  $ is always zero. Then the pdf of $\left(  \ref{PRoabvas copy(1)}%
\right)  $ reads%
\begin{equation}
\widetilde{f}_{\mathbf{X}}=c_{\varnothing}f_{\mathbf{X}}+c_{\left\{
1\right\}  }\widetilde{f}_{\left\{  1\right\}  }+c_{\left\{  2\right\}
}\widetilde{f}_{\left\{  2\right\}  }+c_{\left\{  1,2\right\}  }\widetilde
{f}_{\left\{  1,2\right\}  }\text{.\label{PRoabvas copy(4)}}%
\end{equation}

In general, we start from a set of $L$ views with $L$ potentially different
confidences%
\begin{equation}%
\begin{tabular}
[c]{ccc}%
index & view & confidence\\
$1$ & $\ldots$ & $c_{1}$\\
$2$ & $\ldots$ & $c_{2}$\\
$\vdots$ & $\vdots$ & $\vdots$\\
$L$ & $\ldots$ & $c_{L}$%
\end{tabular}
\end{equation}
From this, we obtain a probability $c_{A}$ for each subset $A$ of $\left\{
1,2,\ldots,L\right\}  $ as follows:%
\begin{align}
\left\{  1,2,\ldots,L\right\}   &  \mapsto c_{\left\{  1,2,\ldots L\right\}
}\equiv\min\left(  c_{l}|l\in\left\{  1,2,\ldots,L\right\}  \right)
\nonumber\\
\left\{  1,2,\ldots,L-1\right\}   &  \mapsto c_{\left\{  1,2,\ldots
,L-1\right\}  }\equiv\min\left(  c_{l}|l\in\left\{  1,2,\ldots,L-1\right\}
\right) \nonumber\\
&  -c_{\left\{  1,2,\ldots L\right\}  }\nonumber\\
&  \vdots\label{Cspafjsgf}\\
\left\{  2,\ldots,L\right\}   &  \mapsto c_{\left\{  2,\ldots,L\right\}
}\equiv\min\left(  c_{l}|l\in\left\{  2,\ldots,L\right\}  \right)
-c_{\left\{  1,2,\ldots L\right\}  }\nonumber\\
\left\{  1,2,\ldots,L-2\right\}   &  \mapsto c_{\left\{  1,2,\ldots
,L-2\right\}  }\equiv\min\left(  c_{l}|l\in\left\{  1,2,\ldots,L-2\right\}
\right) \nonumber\\
&  -c_{\left\{  1,2,\ldots,L-1\right\}  }-c_{\left\{  1,2,\ldots L\right\}
}\nonumber\\
&  \vdots\nonumber\\
\varnothing &  \mapsto c_{\varnothing}\equiv1-\sum_{l=1}^{L}c_{l}\nonumber
\end{align}
The set of subsets is known as the "power set" and is denoted $2^{\left\{
1,\ldots,L\right\}  }$. Therefore, the views and their confidences are mapped
into a probability on the power set of the views.

The posterior is defined in distribution as follows%
\begin{equation}
\widetilde{\mathbf{X}}\overset{d}{=}\sum_{A\in2^{\left\{  1,\ldots,L\right\}
}}I_{c_{A}}\left(  U\right)  \widetilde{\mathbf{X}}_{A}%
\text{,\label{PRoabvas copy(6)}}%
\end{equation}
where $U$ is a uniform random variable; the $I_{c}$'s are indicators functions
of the on non-overlapping intervals with size $c_{A}$ as in $\left(
\ref{Cspafjsgf}\right)  $; the $\widetilde{\mathbf{X}}_{A}$'s are independent
random variables, distributed according to the posterior with only the views
in the set $A$, whose pdf we denote by $\widetilde{f}_{A}$.

The pdf of the posterior $\left(  \ref{PRoabvas copy(6)}\right)  $ then reads%
\begin{equation}
\widetilde{f}_{\mathbf{X}}=\sum_{A\in2^{\left\{  1,\ldots,L\right\}  }}%
c_{A}\widetilde{f}_{A}\text{.\label{PRoabvas copy(7)}}%
\end{equation}
Notice that in practice the vast majority of the potentially $2^{L}$ subsets
will have null probability $c_{A}$ and therefore those terms will not appear
in $\left(  \ref{PRoabvas copy(6)}\right)  $ or $\left(
\ref{PRoabvas copy(7)}\right)  $.

\subsubsection{Multiple users, equal confidence levels in their views}

This is the case considered in the pooling expression $\left(  \ref{Mulaigte}%
\right)  $, which we report here%
\begin{equation}
\widetilde{f}_{\mathbf{X}}^{\mathbf{c}}\equiv\sum_{s=0}^{S}\widetilde{c}%
_{s}\widetilde{f}_{\mathbf{X}}^{\left(  s\right)  }%
\text{.\label{Mulaigte copy(1)}}%
\end{equation}

\subsubsection{Multiple users, different confidence levels in their views
\label{SecMsiaog}}

More in general, consider $S$ users. The generic $s$-th user has $L_{s}$ views
with potentially different relative confidences, modeled as in $\ \left(
\ref{PRoabvas copy(7)}\right)  $. On the other hand, each user has been given
an overall confidence level as in $\left(  \ref{Mulaigte copy(1)}\right)  $.
The pdf of the posterior follows from integrating the bottom-up approach
$\left(  \ref{PRoabvas copy(7)}\right)  $ and the top-down approach $\left(
\ref{Mulaigte copy(1)}\right)  $ as follows:%
\begin{equation}
\widetilde{f}_{\mathbf{X}}=\sum_{s=0}^{S}\widetilde{c}_{s}\sum_{A_{s}%
\in2^{\left\{  1,\ldots,L_{s}\right\}  }}c_{A_{s}}\widetilde{f}_{A_{s}%
}\text{.\label{PRoabvas copy(10)}}%
\end{equation}
We remark that in practice the vast majority of the potentially large number
of the terms $c_{A_{s}}$ in $\left(  \ref{PRoabvas copy(10)}\right)  $ is
null. Also this model can be embedded in the framework of a probability on the
power set of the views, as in $\left(  \ref{PRoabvas copy(6)}\right)
$-$\left(  \ref{PRoabvas copy(7)}\right)  $, see Appendix
\ref{AppGeapfBdvdfio}.

\subsubsection{General case\label{AppGeapfBdvdfio}}

We can interpret the multi-user, multi-confidence framework as a set of
$L\equiv L_{1}+\cdots L_{S}$ views with confidences defined as the product of
the overall confidence in the user times the relative confidence of the user
in his different views.
\begin{equation}%
\begin{array}
[c]{c}%
\text{user }1\text{:}\left\{
\begin{tabular}
[c]{lcc}%
index & view & conf.\\
$\left(  1,1\right)  $ & $\ldots$ & $c_{1,1}$\\
$\left(  1,2\right)  $ & $\ldots$ & $c_{1,2}$\\
$\vdots$ & $\vdots$ & $\vdots$\\
$\left(  1,L_{1}\right)  $ & $\ldots$ & $c_{1,L_{1}}$%
\end{tabular}
\right. \\
\vdots\label{Vasasgfrtgt}\\
\text{user }S\text{:}\left\{
\begin{tabular}
[c]{lcc}%
index & view & conf.\\
$\left(  S,1\right)  $ & $\ldots$ & $c_{S,1}$\\
$\left(  S,2\right)  $ & $\ldots$ & $c_{S,2}$\\
$\vdots$ & $\vdots$ & $\vdots$\\
$\left(  S,L_{S}\right)  $ & $\ldots$ & $c_{S,L_{S}}$%
\end{tabular}
\right.
\end{array}
\end{equation}
Consider the power set%
\begin{equation}
\mathcal{A}\equiv2^{\left\{  \left(  1,1\right)  ,\ldots,\left(
S,L_{S}\right)  \right\}  }\text{.}%
\end{equation}
The sum in $\left(  \ref{PRoabvas copy(10)}\right)  $ can be expressed as
\begin{equation}
\widetilde{f}_{\mathbf{X}}=\sum_{A\in\mathcal{A}}c_{A}\widetilde{f}%
_{A}\text{,\label{PRoabvas copy(2)}}%
\end{equation}
where the coefficients $c_{A}$ are determined by the integration of the
bottom-up approach $\left(  \ref{PRoabvas copy(7)}\right)  $ and the top-down
approach $\left(  \ref{Mulaigte copy(1)}\right)  $:\ due to this integration
only very few among all the possible elements $A\in\mathcal{A}$ have a
non-null coefficient $c_{A}$.

However, there are many choices of the $c_{A}$'s consistent with $\left(
\ref{Vasasgfrtgt}\right)  $. According to any such choice, the posterior is
expressed in distribution as%
\begin{equation}
\widetilde{\mathbf{X}}\overset{d}{=}\sum_{A\in\mathcal{A}}I_{c_{A}}\left(
U\right)  \widetilde{\mathbf{X}}_{A}\text{,\label{PRoabvas copy(8)}}%
\end{equation}
where the same notation as $\left(  \ref{PRoabvas copy(6)}\right)  $ applies,
and the pdf reads%
\begin{equation}
\widetilde{f}_{\mathbf{X}}=\sum_{A\in\mathcal{A}}c_{A}\widetilde{f}%
_{A}\text{.\label{PRoabvas copy(9)}}%
\end{equation}

\end{document}